# PA-ResSeg: A Phase Attention Residual Network for Liver Tumor Segmentation from Multi-phase CT Images

Yingying Xu[1,6], Ming Cai[1,a)], Lanfen Lin[1], Yue Zhang[1], Hongjie Hu[2], Zhiyi Peng[3], Qiaowei Zhang[2], Qingqing Chen[2], Xiongwei Mao[3], Yutaro Iwamoto[4], Xian-Hua Han[5], Yen-Wei Chen[4,6], Ruofeng Tong[1,6]

[1]Collage of Computer Science and Technology, Zhejiang University, Hangzhou, Zhejiang, China

[2]Department of Radiology, Sir Run Run Shaw Hospital, College of Medicine, Zhejiang University, Hangzhou, Zhejiang, China

[3]Department of Radiology, The First Affiliated Hospital, College of Medicine, Zhejiang University, Hangzhou, Zhejiang, China

[4]College of Information Science and Engineering, Ritsumeikan University, Kusatsu, Shiga, Japan

[5]Artificial Intelligence Research Center, Yamaguchi University, Yamaguchi City, Yamaguchi, Japan

[6] Research Center for Healthcare Data Science, Zhejiang Lab, Hangzhou, China

[a)]Author to whom correspondence should be addressed. Electronic mail: cm@zju.edu.cn

## Abstract

**Purpose:** Liver tumor segmentation is a crucial prerequisite for computer aided diagnosis of liver tumors. In the clinical liver tumor diagnosis, radiologists usually examine multi-phase images as these images provide abundant and complementary information of tumors. However, most known automatic segmentation methods extract tumor features from CT images merely of a single phase, in which valuable multi-phase information is ignored. Therefore, it is highly demanded to develop a method effectively incorporating multi-phase information for automatic and accurate liver tumor segmentation.

**Methods:** In this paper, we propose a phase attention residual network (PA-ResSeg) to model multi-phase features for accurate liver tumor segmentation, in which a phase attention (PA) is newly proposed to additionally exploit the images of arterial (ART) phase to facilitate the segmentation of portal venous (PV) phase. The PA block consists of an intra-phase attention (Intra-PA) module and an inter-phase attention (Inter-PA) module to capture channel-wise self-dependencies and cross-phase interdependencies, respectively. Thus it enables the network to learn more representative multi-phase features by refining the PV features according to the channel dependencies and recalibrating the ART features based on the learned interdependencies between phases. We propose a PA-based multi-scale fusion (MSF) architecture to embed the PA blocks in the network at multiple levels along the encoding path to fuse multi-scale features from multi-phase images. Moreover, a 3D boundary-enhanced loss (BE-loss) is proposed for training to make the network more sensitive to boundaries.

**Results:** To evaluate the performance of our proposed PA-ResSeg, we conducted experiments on a multi-phase CT dataset of focal liver lesions (MPCT-FLLs). Experimental results show the effectiveness of the proposed method by achieving a dice per case (DPC) of 0.77.87, a dice global (DG) of 0.8682, a volumetric overlap error (VOE) of 0.3328 and a relative volume difference (RVD) of 0.0443 on the MPCT-FLLs. Furthermore, to validate the effectiveness and robustness of PA-ResSeg, we conducted extra experiments on another multi-phase liver tumor dataset and obtained a DPC of 0.8290, a DG of 0.9132, a VOE of 0.2637 and a RVD of 0.0163. The proposed method shows its robustness and generalization capability in different datasets and different backbones.

**Conclusions:** The study demonstrates that our method can effectively model information from multi-phase CT images to segment liver tumors and outperforms other state-of-the-art methods. The PA-based MSF method can learn more representative multi-phase features at multiple scales and thereby improve the segmentation performance. Besides, the proposed 3D BE-loss is conducive to tumor boundary segmentation by enforcing the network focus on boundary regions and marginal slices. Experimental results evaluated by quantitative metrics demonstrate the superiority of our PA-ResSeg over the best-known methods.

## 1. INTRODUCTION

Hepatic disease is one of the most common diseases in the world, in which malignant liver tumors cause massive deaths every year[1]. Computed Tomography (CT) is one of the most important imaging modalities and is widely applied in diagnosis of liver tumors. Accurate segmentation of tumors from CT images is crucial to computer-aided diagnosis (CAD) of liver tumors. Traditionally, the delineation of liver and liver lesions in the CT images is manually performed by radiologists, which is time-consuming and suffers from inter- and intra-annotator variants. Therefore, it is highly desired to develop an automatic and accurate method for liver and liver tumor segmentation.

Automatic segmentation of liver and liver tumor remains a challenging task due to the size and shape variations of tumors, inter-tumor heterogeneity and low intensity contrast among tumors, liver and other surrounding tissues. Quite a few segmentation methods have been proposed to tackle these difficulties, including hand-crafted feature based methods[2-5] and deep learning based methods[6-8]. Among hand-crafted feature based methods, machine-learning based methods are the most successful and popular techniques for liver and tumor segmentation. However, the hand-crafted feature based methods are prone to achieve limited performance because of the limited representation capability. Recently, deep learning has shown its superiority in various challenging tasks in the medical field, including classification[9,10], detection[11,12] and segmentation[13,14] of medical images. In particular, convolutional neural network (CNN) has become a popular choice for medical image analysis[15,16]. Many fully convolutional network (FCN)-based methods for automatic liver or tumor segmentation have been proposed[8,17,18].

According to the acquiring time before and after contrast injection, a dynamic series of CT images including noncontrast-enhanced (NC) phase, arterial (ART) phase, portal venous (PV) phase and delay (DL) phase can be obtained. Figure 1 provides the imaging appearance of five typical types of liver tumors over multiple phases, i.e., cysts, focal nodular hyperplasia (FNH), hemangiomas (HEM), hepatocellular carcinoma (HCC) and metastasis (METS). The inter-phase characteristic from multi-phase CT images (such as the enhancement pattern) is an essential factor for radiologists to differentiate various focal liver lesions (FLLs). In the proceeding of manual annotation for tumors, multi-phase scans are able to provide complementary information for identifying boundaries especially when the lesion appears to be unclear. For most cases, ART phase may lead to fuzzy boundaries due to simultaneous enhancement of tumors and liver parenchyma, while the maximum contrast between tumors and surrounding liver tissue will occur in the PV phase. Thus, PV phase is preferred for liver tumor segmentation. Note that not all types of tumors are noticeable in the PV phase. Specifically, Figure 2 shows such examples (from the in-house MPCT-FLLs dataset) where tumors are unequivocal in certain phase while ambiguous in another phase. Using only single-phase images may neglect the pivotal and implicit information conveyed by multi-phase scans and limit segmentation accuracy.

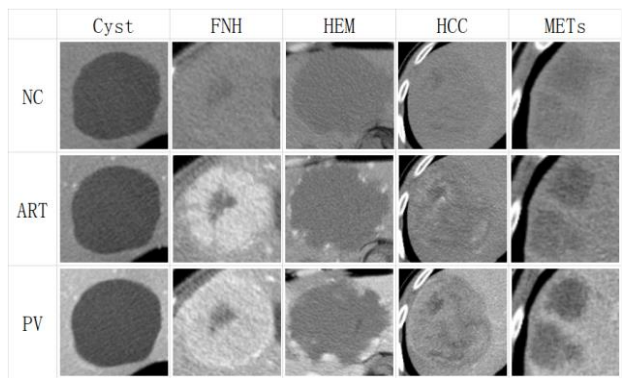

Figure 1. The imaging appearance of five typical types of liver tumors over multiple phases. DL phase is not provided because scans of DL are not collected for benign tumors in most cases in our dataset.

Most of the recently published methods[8,18,19] for liver tumor segmentation are implemented on two public datasets (2017 ISBI LiTS dataset[20] and 3DIRCAD[21]). The public datasets of liver tumor only provide single-phase images of PV phase. The study of liver tumor segmentation based on multi-phase CT images is restricted by dataset acquisition. Currently, several methods[7,22,23] have been proposed, incorporating multi-phase information to improve liver tumor segmentation performance. The general methods for modeling multi-phase images basically include the following two modes: (1) Early fusion that multi-phase images are stacked together as input. (2) Late fusion where independent networks are designed for each phase and the features of output-layer are fused to get the segmentation (such as multi-channel FCNs[7]). A single network for feature extraction of multi-phase images using early fusion may lead to lower representation power for each phase, while independent processing of each phase may

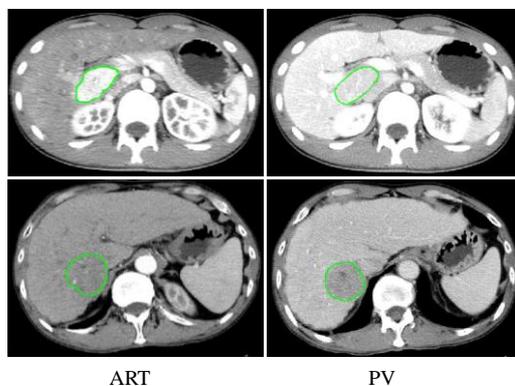

ART           PV

Figure 2. Examples of tumors in ART and PV phases. The contour of the tumor is outlined in green line.

neglect the underlying correlation information among phases that might benefit the segmentation. Both of two fusion modes have no interaction among phases during feature extraction, thus the inter-phase complementary information is not adequately used.

To address the above issues, we propose a phase attention 2D residual network (PA-ResSeg) to fully leverage the multi-phase CT images for accurate liver tumor segmentation. Specifically, we employed both the ART and PV images to take advantage of complementary information from multiple phases to facilitate the segmentation of PV phase. A multi-scale fusion (MSF) strategy is proposed to probe multi-scale features where the multi-phase features are fused on several fusion blocks at multiple levels along the encoding path. Phase attention (PA) is designed to extract much more representative multi-phase features. The PA block consists of an intra-phase attention (Intra-PA) module and an inter-phase attention (Inter-PA) module. The Intra-PA module models channel-wise self-dependencies between channel maps of the PV phase and enhances meaningful features to improve feature representation capability. The Inter-PA module allows to capture the inter-phase channel dependencies between the PV and ART phases and emphasize the features of ART phase that have high correlation with PV phase. The PA-based MSF architecture is finally exploited in our method, where the PA blocks are embedded into the network in multi-level fusion blocks of the MSF structure. To the best of our knowledge, it is the first try to apply attention mechanism in dealing with multi-phase features for liver tumor segmentation. In addition, we propose a 3D boundary-enhanced loss (BE-loss) to make the network to be more sensitive to tumor boundaries and account more 3D spatial contextual information, and then further improve the segmentation performance.

The main contributions of the proposed method are summarized as four-fold:

(1) We develop a PA-ResSeg network to additionally exploit the information carried by images from the ART phase to facilitate the tumor segmentation of the PV phase and thus cope with the situation where different tumors show variable radiological manifestations in different phases. It is worth noting that one single PA-ResSeg model is competent for various types of tumors and all types of tumors are regarded as a single class in our segmentation task without using extra class label.

(2) We devise a PA module to capture the intra-phase dependencies and cross-phase interdependencies, which allows the network to learn representative features of the PV phase based on self-dependencies of channel maps and adaptively select the ART features according to the learned correlation coefficients between the PV and ART phases.

(3) We propose a PA-based MSF architecture to fully leverage the features from multiple phases by integrating multi-scale features from multiple phases in PA blocks, thus the multi-phase features can be jointly and effectively learned.

(4) The proposed 3D boundary-enhanced cross-entropy loss can boost the segmentation on tumor boundaries and further contribute to the performance of tumor segmentation.

## 2. RELATED WORK
### 2.A. Hand-crafted feature based methods

Traditional segmentation methods for liver and liver tumor based on hand-crafted features includes thresholding[2,24], region growing[25], level set[3,26-28], graph-based methods[29-31] and machine learning based methods[4,5,32]. Thresholding based algorithms determine the foreground and background by comparing the gray value of all pixels with one or several thresholds. It is suitable for images where the foreground and background occupy different grayscale ranges. Wong et al.[25] proposed a semi-automated method to segment tumors from 2D slices using 2D region growing with knowledge-based constraints to constrain the size and shape of the segmented region. Level set techniques first proposed by Osher and Sethian[33] have been widely adopted for liver tumor segmentation due to their ability to handle intensity heterogeneities and image noise[26-28]. For example, Li et al.[3] proposed a level set model incorporating likelihood energy and edge energy. The model enabled a better delineation of density distribution of liver tumors and representation of boundaries. Hoogi et al.[27] proposed the adaptive local window to improve level set technique. Graph based methods have gained great attention for medical image segmentation including graph cut[29] and random work[30,31]. Machine-learning based methods especially classification based and clustering based methods are the most common techniques in the image segmentation. Fuzzy C-Means (FCM)[34] is a popularly applied clustering based method. Anter et al.[4] proposed to extract liver from CT images first and then use fast FCM clustering algorithm to segment tumors from segmented livers. Representative classification based methods include Support Vector Machine (SVM)[5] and Random forest[32]. Vorontsov et al.[5] trained a two-class kernel SVM classifier to generate voxel classification map and combined the map and a deformable surface model to perform the segmentation of tumors. Foruzan et al[35] proposed a sigmoid edge model, which was combined with the SVM and watershed algorithms, to improve the lesion segmentation performance. However, these methods yield limited performance due to their low capability of feature representation, especially when coping with complex segmentation scenarios where tumors have low contrast against background tissues.

## 2.B. Deep learning based methods

Recently, deep learning with CNNs has gained impressive performance in medical image analysis[9,11,13]. Fully convolutional neural networks (FCNs) are reported to be popularly applied in liver and liver tumor segmentation[6-8,18,36]. For example, Christ et al.[6] proposed a cascaded FCN (CFCN) that segmented liver and tumors sequentially. The segmented liver ROIs are fed into the second FCN to segment tumors sorely, then dense 3D conditional random fields (3DCRFs) are applied to refine the segmentations. The CFCN was extended to new clinical CT datasets and different modalities in their later work[36]. Li et al.[18] proposed a hybrid densely connected UNet (H-DenseUNet), which is composed of both 2D and 3D denseUNet for intra-slice and 3D contextual feature extraction respectively. Seo et al.[8] added a residual path and additional convolution operations to the skip connection of the U-Net[37] to modify the conventional U-Net and the modified U-Net outperforms other methods on the task of liver and tumor segmentation. Wang et al.[19] introduced a volumetric attention module for 3D liver and tumor segmentation and also evaluated their method in a detection task. Most of the mentioned methods are implemented on single-phase CT images. Some methods based on multi-phase CT images have been reported utilizing an in-house multi-phase CT dataset. Sun et al.[7] proposed a multi-channel fully convolutional network (MC-FCN) with three independent FCN channels for three phases, respectively. The multi-phase features extracted from three channels are fused in the high-level layer to improve the segmentation accuracy. The effectiveness of multi-phase information was also validated in Ouhmich's work[23]. Two different fusion strategies were studied in the work: Dimensional MultiPhase (DMP) and MultiPhase Fusion (MPF). The DMP deals with the NC, ART and PV slices as three channels of the input. As for the MPF, multi-phase images are processed separately and then the output maps are merged (by simple addition) to get the final segmentation. For the above work, the single branch or channel of the network for each phase is trained independently. The inter-phase complementary information cannot be fully mined through independent training. To address this issue, the modality weighted U-Net (MW-UNet)[22] proposed to fuse multi-phase features at several specific layers of the U-Net by weighting the multiple feature maps. The weights that represent the relative importance of the phases are dynamically learned through training.

# 3. MATERIALS AND METHODS

## 3.A. Datasets and Pre-processing

We evaluated our method on an in-house MPCT-FLLs (Multi-Phase CT dataset of Focal Liver Lesions) database retrospectively collected by Sir Run Run Shaw Hospital. A subset of the dataset was already used in our previous work for medical image retrieval[38]. The MPCT-FLLs database contains 121 sets of multi-phase abdominal CT images collected using multidetector helical CT scanners of GE with a slice thickness of 5.0 mm and scanners of SIEMENS with thickness of 7.0 mm. All images are in a slice-plane resolution of 512×512 pixels, with a largely varying in-plane resolution from 0.52 mm×0.52 mm to 0.86 mm×0.86 mm. CT scans of ART and PV phase are selected in our study while NC phase is omitted because of its low inter-tissue contrast and low contribution to the segmentation performance. The number of slices per phase for each case has a range from 25 to 99. The MPCT-FLLs contains 3170 slices for each phase and 6340 (3170×2) slices in total for two phases. The annotation of liver and lesions are manually delineated by two experienced radiologists and examined by another two experienced radiologists as ground truth for segmentation. Five types of common liver tumors with confirmed pathology/diagnosis are collected. Table 1 displays the detailed distribution of different types of tumors and multiple scanners of MPCT-FLLs dataset.

Table 1. The detailed distribution of different types of tumors and multiple scanners in MPCT-FLL dataset: #Cases (number of cases) and #Slices (number of slices)

| Type | Tumor Type | | | | | Scanner | | Total |
|------|------|------|------|------|------|------|---------|-------|
|  | Cyst | FNH | HCC | HEM | METs | GE | SIEMENS |  |
| #Cases | 36 | 20 | 25 | 26 | 14 | 75 | 46 | 121 |
| #Slices | 927×2 | 506×2 | 674×2 | 693×2 | 370×2 | 2179×2 | 991×2 | 3170×2 |

In order to validate the effectiveness and robustness of our proposed PA method and MSF strategy for modelling multi-phase information, we conduct extra experiments on another liver tumor dataset MPCT-HCC (from the Department of Radiology, The First Affiliated Hospital, College of Medicine, Zhejiang University) that also contains multi-phase CT images (ART and PV phases). The MPCT-HCC dataset is consisted of 209 sets of HCC cases collected using a CT scanner of Philips iCT 256. In this dataset, each phase of case consists of slices ranging from 35 to 88 with a slice-plane resolution of 512×512 pixels and a slice thickness of 5.0 mm. The in-plane resolution is in the range of 0.56 mm×0.56 mm and 0.85 mm×0.85 mm. The annotation of lesions are provided by two resident doctors with more than 3-year experience and checked by an experienced radiologist with more than 10-year experience. Among the 209 sets of data, 42 cases are randomly selected for testing and the remaining 167 cases are used for training. Data partitioning is conducted twice for repeated holdout evaluation experiments. The detailed data distribution (number of cases and number of slices) of the MPCT-HCC dataset for two partitions is shown in Table 2.

Figure 3 displays the examples (CT slice of PV phase) of five types of tumors from the MPCT-FLLs dataset and two HCC examples from the MPCT-HCC dataset.

Table 2. The detailed data distribution (numbers of cases and slices) of MPCT-HCC dataset

| Type | Training Set | | Test set | | Total |
|---|---|---|---|---|---|
| Partition | Set1 | Set2 | Set1 | Set2 | |
| #Cases | 167 | 167 | 42 | 42 | 209 |
| #Slices | 7377×2 | 7472×2 | 1922×2 | 1827×2 | 9299×2 |

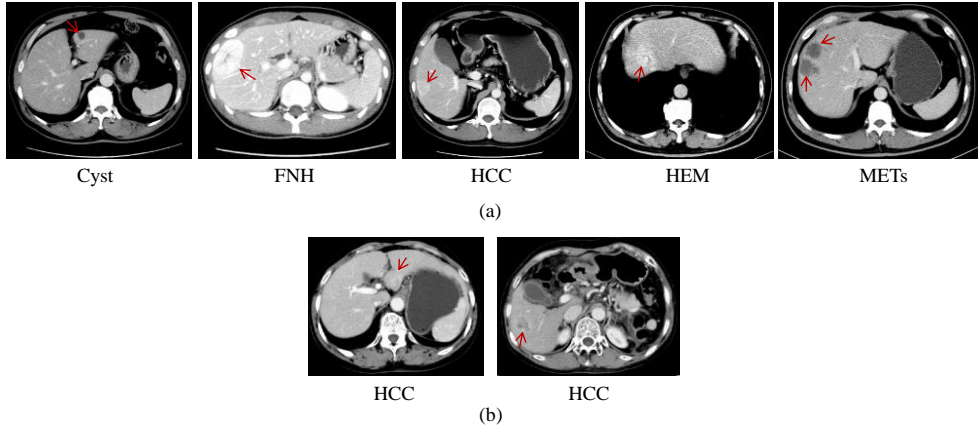

Figure 3. Examples (CT slice of PV phase) from two in-house dataset. (a) five types of tumors from the MPCT-FLLs dataset. (b) two HCC examples from the MPCT-HCC dataset. The lesions are indicated by the red arrow.

For image preprocessing, we truncate the image intensity values to the range of [-70 HU, +180 HU] according to the commonly used window width/window level for abdominal CT images in clinical practice [39]. During a clinical CT study, spatial aberration in multi-phase images is inevitable due to difference in patient body position, respiratory movements and heartbeat. A non-rigid registration algorithm based on B-spline[40] is employed in a simple but effective way to align the images of the ART phase with images of the PV phase to overcome the spatial aberration in some extent.

## 3.B. Network Architecture Overview

In this subsection, we first give an overview of our proposed method (PA-ResSeg) for liver tumor segmentation based on multi-phase CT images. As illustrated in Figure 4, the general framework of the network appears to be a multi-branch architecture with two interactive phase-specific encoders and a shared decoder. Features from multiple phases are extracted through multiple encoders and the phase attention (PA) is designed to model cross-phase features. The PA blocks are integrated to the multi-branch network at multiple levels along the encoding path using the multi-scale fusion (MSF) strategy to make full mining of multi-scale features from multi-phase images. The multi-scale output feature maps of the PA blocks are gradually up-sampled to the original size to get the final output. To achieve a more precise segmentation of boundaries, a 3D boundary-enhanced loss (BE-loss) is used to supervise the training of the segmentation network. The elaboration of the MSF, PA block and 3D BE-loss will be presented in the following subsections.

We employ a 2D residual network as the backbone for encoder of each branch. Inspired by the work of VoxResNet[11] which was proposed for 3D brain segmentation, we adopt the upsampling strategy by gradually upsampling multi-level feature maps, which is different from the symmetrical architecture of the UNet[37]. The basic single-branch residual segmentation network is named ResSeg. In the ResSeg network, the multi-scale output maps of convolution layers or residual blocks are gradually up-sampled to get the output maps with the same resolution as input, where both low-level spatial features and high-level semantic features are preserved for better segmentation. Each up-sampling operation is followed by a batch normalization[42] and a ReLU activation function[43]. The multiple output maps are concatenated to get the final probability map. The detailed structure of the ResSeg is described in Table 3. We also use the UNet as the backbone for comparison in our experiments and the results show that the proposed ResSeg outperforms the UNet, which will be elaborated in Section 4.

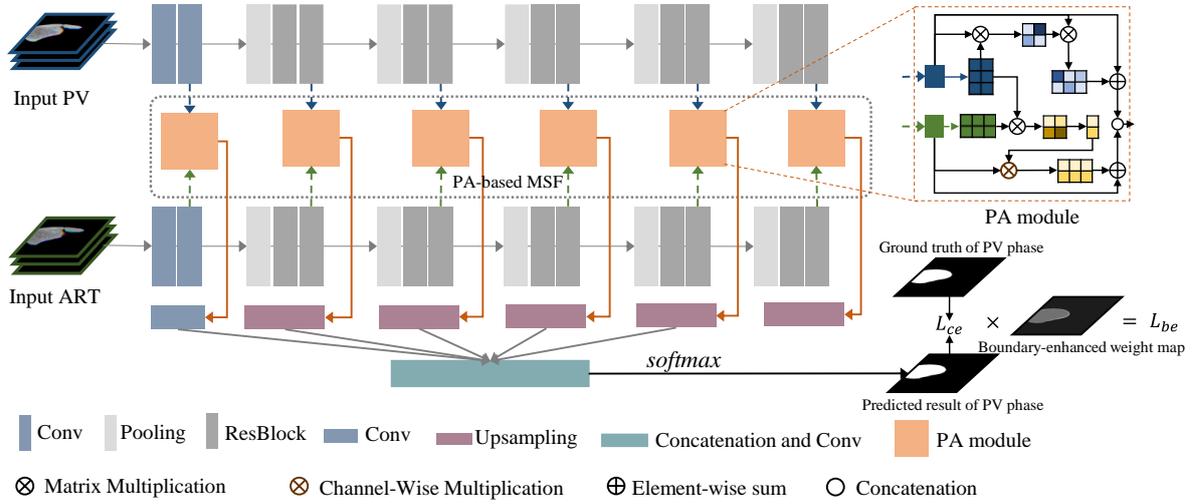

Figure 4. An illustration of the proposed multi-channel PA-ResSeg

Table 3. The detailed operation and parameters of each layer in the ResSeg.

| Encoder | | | Decoder | |
|---|---|---|---|---|
| Layer name | Output size | Operation, kernel size, out channels, stride | Layer name | Operation, kernel size, out channels, stride, output size |
| conv1_1 | 224×224 | conv, 3 × 3, 64, stride 1 | conv1_3 | conv, 3×3, 16, 1, 224 |
| conv1_2 | 224×224 | conv, 3 × 3, 64, stride 1 | | |
| conv2_x | 112×112 | max pool, 2 × 2, stride 2 | upsampling_1 | deconv, 4×4, 16, 2, 224 |
| | | $\begin{bmatrix} \text{conv, 3×3, 128, stride 1} \\ \text{conv, 3×3, 128, stride 1} \end{bmatrix} \times 2$ | | |
| conv3_x | 56×56 | max pool, 2 × 2, stride 2 | upsampling_2 | deconv, 4×4, 32, 2, 112 |
| | | $\begin{bmatrix} \text{conv, 3×3, 256, stride 1} \\ \text{conv, 3×3, 256, stride 1} \end{bmatrix} \times 2$ | | deconv, 4×4, 16, 2, 224 |
| conv4_x | 28×28 | max pool, 2 × 2, stride 2 | upsampling_3 | deconv, 4×4, 64, 2, 56 |
| | | $\begin{bmatrix} \text{conv, 3×3, 512, stride 1} \\ \text{conv, 3×3, 512, stride 1} \end{bmatrix} \times 2$ | | deconv, 4×4, 32, 2, 112 |
| | | | | deconv, 4×4, 16, 2, 224 |
| conv5_x | 14×14 | max pool, 2 × 2, stride 2 | upsampling_4 | deconv, 4×4, 128, 2, 28 |

This is a self-archival version from the author.

| | | | | deconv, 4×4, 64, 2, 56 |
|---|---|---|---|---|
| | | $\left[\begin{array}{l}\text{conv, 3×3, 512, stride 1}\\\text{conv, 3×3, 512, stride 1}\end{array}\right]\times 2$ | | deconv, 4×4, 32, 2, 112 |
| | | | | deconv, 4×4, 16, 2, 224 |
| conv6_x | 7×7 | max pool, 2 × 2, stride 2 | upsampling_5 | deconv, 4×4, 256, 2, 14 |
| | | $\left[\begin{array}{l}\text{conv, 3×3, 512, stride 1}\\\text{conv, 3×3, 512, stride 1}\end{array}\right]\times 2$ | | deconv, 4×4, 128, 2, 28 |
| | | | | deconv, 4×4, 64, 2, 56 |
| | | | | deconv, 4×4, 32, 2, 112 |
| | | | | deconv, 4×4, 16, 2, 224 |
| | | | final | concat, 224 |
| | | | | conv, 3×3, 96, stride 1, 224 |
| | | | | conv, 3×3, 2, stride 1, 224 |
| | | | | softmax |

In our experiments, we first train a simple ResSeg (named as liver-ResSeg) on the training data of the MPCT-FLLs dataset (including both ART and PV images) for coarse liver segmentation to get the liver region of interest (ROI) for both ART and PV phases. Morphological operation of dilation with a 5×5 kernel is performed on the preliminarily segmented livers. Then the value of pixels outside the liver ROIs are set as zero and then fed the processed image into the PA-ResSeg for tumor segmentation.

### 3.C Multi-Scale Fusion (MSF) strategy

The fusion method of multiple data is a matter of concern when dealing with multi-model or multi-phase images. Ouhmich[23] introduced two strategies for incorporating multi-phase images: DMP (Dimensional MultiPhase) which concatenates multi-phase images in the input and MPF (MultiPhase Fusion) which adds the multi-phase output maps where the output maps are computed independently for each phase. The DMP strategy fuses the multiple images in the input while the MFP strategy extracts multi-phase features independently. Both strategies may not incorporate inter-phase complementary information efficiently due to the lack of interaction among different phases during feature extraction. To address this issue, we propose a multi-scale fusion (MSF) strategy to probe multi-scale inter-phase features. In the MSF architecture, a multi-branch network is designed to extract multi-phase information. Different from the MPF, the multi-phase features are fused on several fusion blocks at multiple levels along the encoding path and the fused multi-scale features are then propagated to a shared decoder for the final segmentation. The multi-scale features from multiple phases are extracted and jointly learned through the MSF structure. Figure 5 displays the illustration of the simplified fusion strategies.

### 3.D. Phase Attention Block

In our work, we aim at taking full advantage of multi-phase information from both the ART and PV

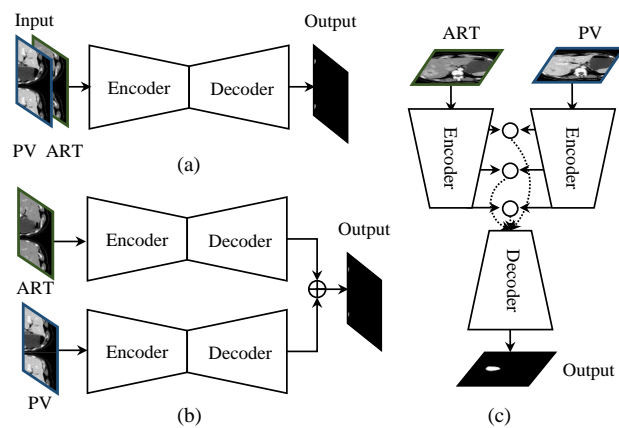

Figure 5. The simplified illustration of three fusion strategies. (a) DMP, (b) MPF, (c) MSF. The circle in the MSF is a fusion block, which can be either concatenation, addition or other operations that fuses the multi-phase features.

phases to boost the segmentation performance of the PV phase. The PA block consists of an intra-phase attention (intra-PA) module and an inter-phase attention (inter-PA) module. We employ the intra-PA module to refine the features of the PV phase based on the channel-wise self-dependencies, and exploit the inter-PA module to recalibrate the ART features according to the cross-phase dependencies between the ART and PV phases. The structure of the PA block is illustrated in Figure 6.

### 3.D.1. Intra-PA module

We implement the intra-PA simply by following the channel attention module proposed by the work of DANet[44]. We first feed the feature map of the PV phase to a convolution layer to get a new feature map $\mathbf{P} \in \mathbb{R}^{C \times H \times W}$. Then the feature map $\mathbf{P}$ is reshaped to $\mathbf{P}' \in \mathbb{R}^{C \times N}$, where $N = H \times W$. The self-channel attention map $\mathbf{M}$ is obtained by a matrix multiplication between $\mathbf{P}'$ and the transpose of $\mathbf{P}'$ (denoted as

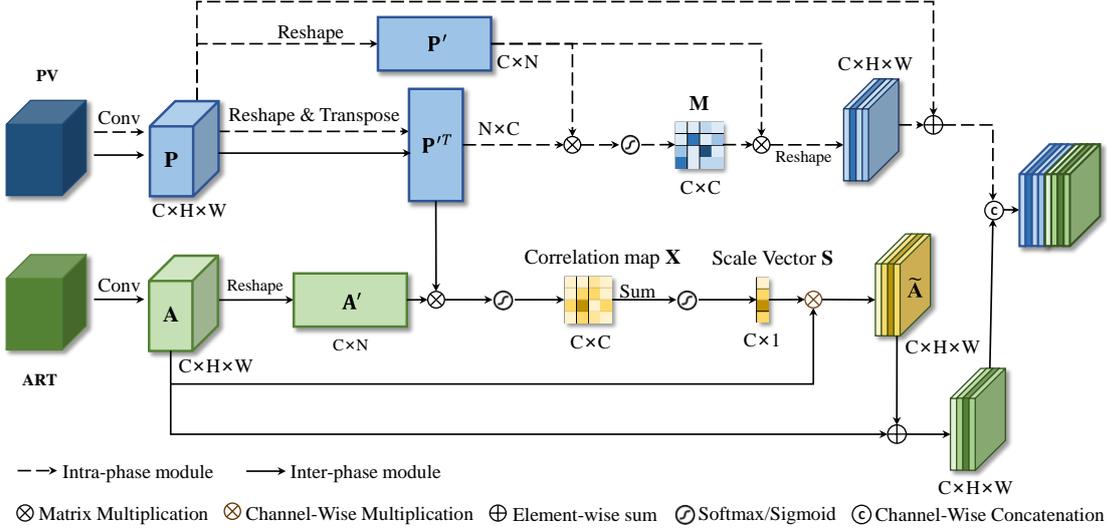

Figure 6. The structure of the PA module

$\mathbf{P}'^T$) followed by a softmax layer:

$$m_{ji} = \frac{exp\ (\mathbf{P}_i' \cdot \mathbf{P}'^T{}_j)}{\sum_{i=1}^{C} exp\ (\mathbf{P}_i' \cdot \mathbf{P}'^T{}_j)} \tag{1}$$

where $m_{ji}$ measures the dependency between the $i^{th}$ channel map and the $j^{th}$ channel map. We perform a matrix multiplication between the transpose of $\mathbf{M}$ and $\mathbf{P}'$ to get the refined feature map and reshape the feature map to $\mathbb{R}^{C \times H \times W}$. An element-wise sum is performed on the refined feature map and the original map $\mathbf{P}$ to get the final output of the intra-PA module. The intra-PA module helps enhance the feature representations of the PV phase based on the self-channel attention.

### 3.D.2. Inter-PA module

As for the inter-PA module, similarly, the feature map of the ART phase is feed to another convolution layer to generate a new feature map $\mathbf{A} = [\boldsymbol{a_1}, \boldsymbol{a_2}, ..., \boldsymbol{a_C}]^T$, where $\mathbf{A} \in \mathbb{R}^{C \times H \times W}$ and $\boldsymbol{a_c} \in \mathbb{R}^{H \times W}$ is the $c^{th}$ channel map of $\mathbf{A}$. Then we reshape $\mathbf{A}$ to get a feature map $\mathbf{A}' \in \mathbb{R}^{C \times N}$. After that a matrix multiplication is performed between $\mathbf{A}$ and $\mathbf{P}'^T$ followed by a softmax layer to get the correlation coefficient map $\mathbf{X} \in \mathbb{R}^{C \times C}$:

$$x_{ij} = \frac{exp\ (\mathbf{A}_i' \cdot \mathbf{P}'^T{}_j)}{\sum_{i=1}^{C} exp\ (\mathbf{A}_i' \cdot \mathbf{P}'^T{}_j)} \tag{2}$$

where $x_{ij}$ measures the correlation coefficient of the $j^{th}$ channel of PV and the $i^{th}$ channel of ART.

Each row vector $\boldsymbol{x_c} \in \mathbb{R}^{1 \times C}$ of the correlation matrix $\mathbf{X} = [\boldsymbol{x_1}, \boldsymbol{x_2}, ..., \boldsymbol{x_C}]^T$ represents the dependencies between each channel map in the ART phase and all channel maps in the PV phase. We obtain the channel scale vector $\mathbf{S} = [s_1, \ s_2, ..., s_C \ ]^T$ by applying a simple sum operation on the correlation map $\mathbf{X}$ with a sigmoid activation:

$$s_c = \frac{1}{1+exp(\sum_j x_{cj})} \tag{3}$$

$s_c$ is the total correlation of the $c^{th}$ channel of ART and PV features. We rescale the channel maps of ART by performing a channel-wise multiplication between the original feature map $\mathbf{A}$ and the scale vector $\mathbf{S}$. Then an element-wise sum operation is performed on the multiplied matrix and original ART feature map $\mathbf{A} = [\boldsymbol{a_1}, \boldsymbol{a_2}, ..., \boldsymbol{a_C}]^T$ to obtain the final representation $\widetilde{\mathbf{A}} \in \mathbb{R}^{C \times H \times W}$ of inter-PA module.

$$\widetilde{\boldsymbol{a}}_c = F_{rescale}(\boldsymbol{a}_c, s_c) + \boldsymbol{a}_c \tag{4}$$

where $\widetilde{\mathbf{A}} = [\widetilde{\boldsymbol{a}}_1, \widetilde{\boldsymbol{a}}_2, ..., \widetilde{\boldsymbol{a}}_C]^T$ and $\{\boldsymbol{a}_c, \widetilde{\boldsymbol{a}}_c\} \in \mathbb{R}^{H \times W}$. $F_{rescale}(\boldsymbol{a}_c, s_c)$ in Eq. (4) refers to the multiplication between $\boldsymbol{a}_c$ and the scalar $s_c$. $\boldsymbol{a}_c$ is the $c^{th}$ channel map of feature map A and $s_c$ is the $c^{th}$ value of the scale vector $\mathbf{S}$ corresponding to the weight of the $c^{th}$ channel.

By exploiting the inter-PA, we could emphasize the channel maps in the ART phase that have high correlation with the PV phase and suppress the features in the ART phase that have little impact in boosting the segmentation of the PV phase. The refined feature maps of the intra-PA module and the inter-PA module are concatenated along the channel dimension (channel-wise concatenation) to get the final output of the PA block.

### 3.E. PA-based MSF

We propose a PA-based MSF architecture to employ the PA block in the MSF architecture to recalibrate multi-phase features at multiple scales instead of directly concatenating or adding the multi-phase features. As illustrated in Figure 4, we integrate the PA blocks in several specific layers at multiple levels along the encoding path. Multi-scale features of multiple phases are adaptively selected and fused by the PA block.

### 3.F. 3D boundary-enhanced loss

We have observed that the boundaries of the tumor in a slice and the marginal slices along the z-axis are prone to be poorly segmented. To make the network more sensitive to the boundaries and account more contextual information, a 3D boundary-enhanced loss (3D BE-loss) is proposed.

To that end, we initially compute a boundary-enhanced weight map W according to the ground truth mask. Then, a weighted binary cross-entropy loss, i.e. 3D BE-loss, is obtained via multiplying binary cross-entropy loss by W. To be more specific, a 3×3×3 mean filter is first used on the 3D ground truth mask volume to generate a soft 3D mask V. Then a linear transformation is conducted on the soft masks to increase the weight of boundaries and marginal slices. W is finally computed using Eq. (5), in which different transformation functions are used for tumor pixels and background pixels:

$$w_p = \begin{cases} -\alpha v_p + \alpha + W_1, & if \ y_p = 1 \\ \beta v_p + W_2, & otherwise \end{cases} \tag{5}$$

where $W_1$ and $W_2$ are the class weights of the tumor and background predefined as 0.8 and 0.2 for alleviating the problem of class imbalance. p represents a pixel. $y_p$ is the ground truth label of the pixel p and $y_p$ is equal to 1 when the p belongs to tumor. $\alpha$ and $\beta$ are transform coefficients that are set to 1 and 0.5 respectively in our experiments. $v_p$ is the value of the soft mask map V.

The final BE-loss is defined as:

$$L_{BE} = \frac{1}{N} \sum_p -w_p \cdot \left[ y_p \cdot \log(\hat{y}_p) + (1 - y_p) \cdot \log(1 - \hat{y}_p) \right] \tag{6}$$

where $\hat{y}_p$ represents the probability that the pixel p is predicted to be the tumor. $w_p$ is the final transformed weight of the pixel p.

Figure 7 gives a more intuitive illustration of the transformation of the weights for boundaries pixels, background pixels and inner-tumor pixels. We can see from the figure that the weights of pixels in/near the boundaries are increased after the transformation. Similarly, the weights of pixels in marginal slices are increased and greater than those in middle slices along the z-axis.

# 4. EXPERIMENTAL RESULTS

## 4.A. Evaluation Methodology

To quantitatively measure the segmentation performance, seven metrics are used to evaluate the segmentation results, including the dice per case (DPC), dice global (DG), volumetric overlap error

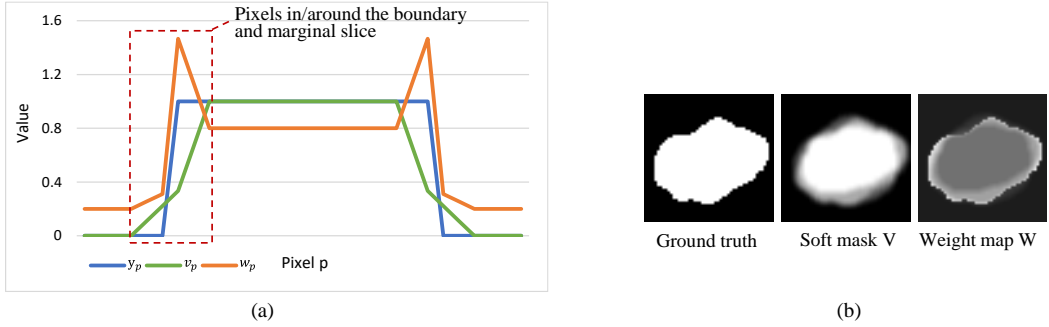

Figure 7 (a) An illustration of the weight transformation. The blue curve is the ground truth label $y_p$ of the pixels. The green curve and the orange curve represent the values of the soft mask $v_p$ and the transformed weight map $w_p$, respectively. (b) An example slice of the ground truth mask, the soft mask V and the weight map W.

(VOE) , relative volume difference (RVD), Sensitivity (TPR), Specificity (TNR) and mean pixel accuracy (ACC). Both of the DPC and DG are calculated based on the DICE score which is formulated as:

$$\text{DICE}(A, B) = \frac{2 \times |A \cap B|}{|A| + |B|} \tag{7}$$

where A is the ground truth and B is the predicted mask. Then DPC and DG score are obtained by averaging DICE score of each case and computing DICE score of whole testing cases, respectively. VOE is the complement of the IoU (Intersection over Union) and is defined as:

$$\text{VOE}(A, B) = 1 - \text{IoU}(A, B) = 1 - \frac{|A \cap B|}{|A \cup B|} \tag{8}$$

RVD measures the relative difference of two volumes and is defined as follows:

$$\text{RVD}(A, B) = \frac{|B| - |A|}{|A|} \tag{9}$$

The greater the DICE score is, the closer the segmentation result is to the ground truth. The VOE and RVD will equal to 0 when the ground truth A and the predicted mask B is the same.

Sensitivity (True Positive Rate, TPR) measures the proportion of positives that are correctly segmented, which is defined as:

$$TPR = \frac{TP}{TP + FN} \tag{10}$$

where TP (true positive) is the number of tumor pixels corrected predicted to be tumor and FN (false negative) is the number of tumor pixels identified as background. Specificity (True Negative Rate, TNR) measures the proportion of negatives that are correctly identified, which is defined as follows:

$$TNR = \frac{TN}{FP + TN} \tag{11}$$

where TN (true negative) is the number of background pixels that are corrected identified as background and FP (false positive) is the number of background pixels that are wrongly recognized as tumors. Mean pixel accuracy (ACC) calculates the average ratio of correctly classified pixels in each class.

We adopted the 5-fold cross-validation technique to estimate models on the MPCT-FLLs dataset and repeated 20%-holdout method on the MPCT-HCC dataset. For the MPCT-FLLs dataset, 121 cases were randomly partitioned into five mutually exclusive subsamples with a similar proportion of different kinds of tumors. The segmentation results are averaged over the 5 test set. For the MPCT-HCC dataset, we conducted experiments twice with 20% data holdout as test data. The results are average values of the two-round experiments.

## 4.B. Implementation and Experimental Setup

In the training phase, we adopted random rotation and flipping for data augmentation to alleviate the problem of overfitting. An input with the size of 224×224×3 was randomly cropped based on the region of liver from three adjacent slices along z-axis in a 2.5D manner during the training. An input pair cropped in the same position from slices of the PV and ART phases were fed into the network. In the test phase, we fed patches (with the size of 224×224×3) cropped from three full slices into the network, and stitched the segmented patches to yield the final prediction of whole slice in an overlap-tile strategy. The segmented results of slices are stacked into a volume to complete the 3D segmentation. The segmentation performance are measured based on the whole volume. We trained our model using a batch size of 8 and utilized the Adam optimizer[455] with a learning rate of 0.0005. The model was trained from scratch and we stop training after 500 epochs without using validation set.

The proposed method was implemented based on the publicly available framework PyTorch 1.0.0[466]. The experiments were carried on a platform with a NVIDIA GTX 1080ti GPU (with 11GB memory) and an six-core PC with an Intel(R) Core(TM) i7-8700K CPU @3.7 GHz (with 32GB RAM).

## 4.C. Ablation study

In this section, to validate the effectiveness of each component in our proposed PA-ResSeg, we conduct an ablation study in the MPCT-FLLs dataset. We start with a single ResSeg and progressively extend it with our MSF architecture, the PA block and the 3D BE-loss. The single ResSeg is trained using only PV images for tumor segmentation of the PV phase. The multi-phase features in the MSF-ResSeg with MSF-only architecture are simply fused by channel-wise concatenation in our experiments. We also construct a single ResSeg with intra-PA module (row 2 in Table 4) and a MSF-ResSeg network with only intra-PA modules for both phases (row 4 in Table 4) to validate the benefit of each module in the PA block. The segmentation results of the ablation study are presented in Table 4. The rows with both Intra-PA module and Inter-PA module checked are architecture with the entire PA block. As we can see from the results, the MSF structure encoding multi-phase features outperforms the single-branch network using only single-phase images. Our PA module helps achieve superior performance compared to the MSF architecture in terms of all the evaluation metrics. The comparison results demonstrate that not only the intra-PA module but also the inter-PA module contribute to the improvement of performance. The 3D BE-loss further contributes to the segmentation performance with 0.61% improvement on DPC for tumor segmentation.

Table 4. Segmentation results by ablation study

| Combination | | | | | Evaluation metrics | | | |
|---|---|---|---|---|---|---|---|---|
| Single | Multi-scale fusion | Intra-PA module | Inter-PA module | BE-loss | DPC (%) | DG (%) | VOE (%) | RVD (%) |
| √ | | | | | 67.87 | 84.31 | 43.18 | 11.23 |
| √ | | √ | | | 70.25 | 83.82 | 40.15 | 19.87 |
| | √ | | | | 74.99 | 86.18 | 36.29 | 5.10 |
| | √ | √ | | | 75.60 | 86.17 | 36.20 | 12.79 |
| | √ | √ | √ | | 77.26 | 86.21 | 33.46 | 4.46 |
| | √ | √ | √ | √ | **77.87** | **86.82** | **33.28** | **4.43** |

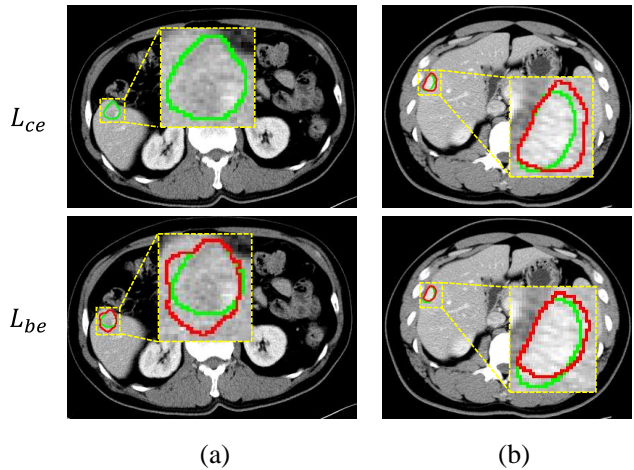

Figure 8. The qualitative results of using different loss. (a) is the segmentation result of a marginal slice and (b) is the segmentation result of a middle slice. The green and red outlines represent ground truth and segmentation results, respectively.

The qualitative results of the models trained with the binary cross-entropy loss ($L_{ce}$) and the 3D BE-loss ($L_{be}$) are shown in Figure 8. Figure 8(a) shows an example of marginal slice where the tumor can be properly segmented by the model trained using BE-loss while CE-loss fails to predict it. We can conclude from the results that the proposed method with BE-loss achieves better results when tackling intra-slice boundary segmentation and marginal slice segmentation.

To validate the generalization capability of MSF strategy and PA block on different backbones, we replace the backbone from ResSeg to UNet[37]. Table 5 reports the quantitative results of different architectures (single UNet using single-phase images of the PV phase, the MSF-UNet fusing the multi-phase features with the MSF strategy and the PA-UNet utilizing the proposed PA-based MSF structure). We can see that the PA-UNet achieved a better performance. The results in Table 4 and Table 5 indicate the effectiveness and robustness of our proposed PA module on both backbones. It is observed that the DG measurement of the single-branch UNet is greater than that of the MSF architecture using multi-phase information. We have analyzed the results and found that the single-branch UNet performs pretty well in the segmentation of Cyst. It might be attributed to the imaging appearance of Cyst with homogeneous density and presenting no enhancement over multiple phases. Thus single-branch UNet is capable to segment Cysts properly and obtain a great DG which may be dominated by the good segmentation results of Cysts, while multi-phase information makes little contribution to performance improvement of Cysts.

Table 5. The results of different architectures using UNet as backbone

| Architecture | DPC (%) | DG (%) | VOE (%) | RVD (%) | TPR (%) | TNR (%) | ACC (%) |
|---|---|---|---|---|---|---|---|
| Single UNet | 68.23 | 85.32 | 41.91 | -12.26 | 66.93 | **99.95** | 83.44 |
| MSF-UNet | 73.63 | 85.21 | 37.73 | 5.16 | 74.89 | 99.93 | 87.42 |
| PA-UNet | **74.92** | **85.39** | **36.38** | **0.90** | **76.12** | 99.94 | **88.02** |

### 4.D. Different fusion strategies

To demonstrate the effectiveness of our strategies for fusing multi-phase information, we compared the proposed PA-ResSeg (MSF with PA) with other three different fusion strategies (DMP[23], MPF[23], MSF without PA (MSF-ResSeg), which are shown in Figure 4, all of which take the ResSeg as segmentation backbone. The strategy of MSF without PA is also our proposed fusion method. The comparison results are shown in Table 6. Note that all methods using multi-phase images surpass the network using only single phase CT scans (shown in Table 4) by a large margin, demonstrating the effectiveness of multi-phase information on liver tumor segmentation. Furthermore, the experiments indicate that the MSF strategy that merges the multi-phase features in several specific layers (no matter without or with PA) gains more improvement comparing to the other two fusion strategies. The MSF with PA achieves the best tumor segmentation results, surpassing the DMP and MPF with 5.37% and 4.02% improvement on DPC respectively.

Table 6. Comparison of different fusion strategies for multi-phase CT images

| Fusion Strategy | DPC (%) | DG (%) | VOE (%) | RVD (%) | TPR (%) | TNR (%) | ACC (%) |
|---|---|---|---|---|---|---|---|
| DMP[23] | 71.89 | 82.87 | 38.97 | 5.73 | 73.38 | 99.93 | 86.65 |
| MPF[23] | 73.24 | 70.60 | **34.38** | 24.93 | **82.10** | 95.08 | **88.59** |
| **MSF w/o PA (MSF-ResSeg)** | **75.00** | **86.18** | 36.29 | **5.10** | 75.98 | **99.95** | 87.96 |
| **MSF with PA (PA-ResSeg)** | **77.26** | **86.22** | **33.46** | **4.46** | **78.82** | 99.94 | **89.38** |

### 4.E. Comparison with State-of-the-arts

We compare our method with the current state-of-the-art methods that are also proposed for liver tumor segmentation based on multi-phase CT images. The MC-FCN is the multi-channel FCNs proposed by Sun et al.[7]. The MW-UNet refers to the modality-weighted UNet introduced by Wu et al.[22]. We implement both methods on our dataset using ART and PV images with the same preprocessing as our method for fair comparison. As only images of ART and PV phase are used in our experiments, the initial phase weights (*β, γ*) of the ART and PV phases in the MW-UNet are set to (0.5, 0.5). The final weights are learned to be (0.348, 0.652), which demonstrates that PV phase contributes more to this segmentation task.

Table 7 displays the segmentation results of different methods and Table 8 reports the performance of the state-of-the-art methods and proposed method on different types of tumors. We can see from the table that our method shows its superiority in the tumor segmentation task compared to the other two methods in terms of seven metrics and performs best on every type of tumor. To further illustrate the superior performance of our method, the qualitative comparisons of the segmentation results are displayed in

| Method | Cyst | FNH | HCC | HEM | METs | Whole test set |
|---|---|---|---|---|---|---|
| MC-FCN7 | 67.82 | 21.58 | 46.98 | 51.59 | 61.07 | 51.80 |
| MW-UNet22 | 83.07 | 64.18 | 64.80 | 71.42 | 79.35 | 73.37 |
| PA-ResSeg | **85.70** | **73.13** | **71.89** | **74.63** | **79.64** | **77.26** |

. It is observed that PA-ResSeg can achieve much better results than the MC-FCN and the MW-UNet.

Table 7. Comparison results of different state-of-the-art methods

| Method | DPC (%) | DG (%) | VOE (%) | RVD (%) | TPR (%) | TNR (%) | ACC (%) |
|---|---|---|---|---|---|---|---|
| MC-FCN[7] | 51.80 | 71.63 | 59.30 | -5.92 | 51.74 | 99.93 | 75.84 |
| MW-UNet[22] | 73.37 | **86.34** | 3710 | 10.11 | 77.17 | 99.93 | 88.55 |
| PA-ResSeg | **77.26** | 86.21 | **33.46** | **4.46** | **78.82** | **99.94** | **89.38** |

Table 8. Segmentation results (DPC: %) of different state-of-the-art methods regarding to different types of tumors.

| Method | Cyst | FNH | HCC | HEM | METs | Whole test set |
|---|---|---|---|---|---|---|
| MC-FCN[7] | 67.82 | 21.58 | 46.98 | 51.59 | 61.07 | 51.80 |
| MW-UNet[22] | 83.07 | 64.18 | 64.80 | 71.42 | 79.35 | 73.37 |
| PA-ResSeg | **85.70** | **73.13** | **71.89** | **74.63** | **79.64** | **77.26** |

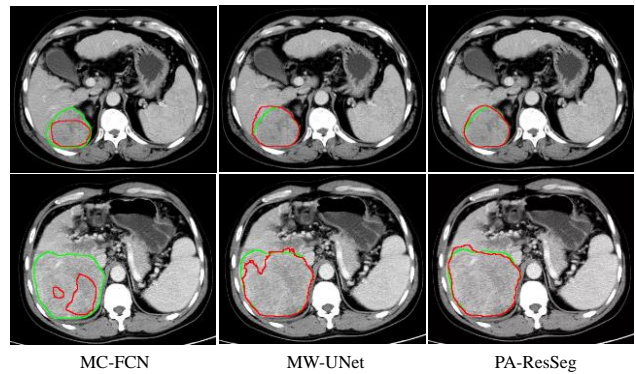

MC-FCN          MW-UNet          PA-ResSeg

Figure 9. Two examples of segmentation results by different methods. The green and red outlines represent ground truth and segmentation results, respectively.

## 4.F. Validation of generalization

In order to validate the generalization capability of the proposed PA-ResSeg for liver tumor segmentation based on multi-phase images, we conduct experiments on data collected from different scanners to explore the generalization of our model on multi-source data. The detailed distribution of data from different scanners is displayed in Table 1. We first train a PA-ResSeg model on the data from GE and then test on the SIEMENS data. Also, a PA-ResSeg model trained on the SIEMENS data is applied on the data from GE. The test results are shown in Table 9. We can see from the table that the model trained on higher resolution data generalize better on low resolution data compared to model trained on low resolution data.

Table 9. Segmentation result of models trained and test on multi-source data

| Train | Test | DPC (%) | DG (%) | VOE (%) | RVD (%) |
|-------|------|---------|--------|---------|---------|
| GE data | SIEMENS data | 71.95 | 82.84 | 39.49 | -0.19 |
| SIEMENS data | GE data | 69.81 | 83.68 | 41.40 | 0.80 |

In addition, to validate the effectiveness and robustness of our proposed PA-ResSeg for liver tumor segmentation based on multi-phase images, we conduct extra experiments on the MPCT-HCC dataset. The same preprocessing is performed on the MPCT-HCC dataset as that on the MPCT-FLLs dataset. Liver segmentation is first performed prior to the tumor segmentation to get the liver ROI using the liver-ResSeg model previously trained on the MPCT-FLLs dataset.

We implement the MSF-ResSeg and PA-ResSeg on the MPCT-HCC dataset and apply the single ResSeg on single-phase images (PV phase) for comparison. The experimental results are shown in Table 10. We can also conclude from the results that using multi-phase information gains greater performance than using single-phase images. The experiments on MPCT-HCC dataset show that the integration of the PA block in the network enables 2.85% improvement of DPC in the segmentation performance compared to the results without the PA block and further demonstrate the effectiveness of our PA method.

Table 10. The segmentation results on MPCT-HCC dataset

| Method | DPC (%) | DG (%) | VOE (%) | RVD (%) | TPR (%) | TNR (%) | ACC (%) |
|--------|---------|--------|---------|---------|---------|---------|---------|
| Single ResSeg | 74.03 | 89.64 | 35.01 | 5.21 | 74.03 | 99.94 | 86.99 |
| MSF-ResSeg | 80.05 | 90.79 | 29.44 | -9.59 | 78.80 | **99.96** | 89.38 |
| **PA-ResSeg** | **82.90** | **91.32** | **26.37** | **1.63** | **83.45** | **99.96** | **91.70** |

## 4. DISCUSSION

Contrast-enhanced CT is one of the most important modality for liver tumor diagnosis. Multi-phase CT images provide abundant and complementary information for liver tumor segmentation. The PV phase is preferred for liver tumor segmentation according to clinical observation. While images of a single phase is incapable to accurately segment all types of tumors because of the situation that different tumors present various clarity in different phase. In this paper, we propose a PA-ResSeg network based on the attention mechanism to additionally exploit the information from the ART phase to facilitate the segmentation of the PV phase. Our method yields a 13.03% improvement of dice per case compared to the single ResSeg using only single-phase images of the PV phase, demonstrating the effectiveness of our PA-ResSeg and the additional utilization of the images from the ART phase.

We implemented our method on a dataset with five types of liver tumors. To intuitively present the impact of our method on different types of tumors, we report the results of Cyst, FNH, HCC, HEM and METs respectively in Table 11 and analyze the effectiveness of our method regarding five types of tumors. We can observed from the results that our method improves the segmentation accuracy on every type of tumors, especially for the FNH which appears to be isodense with the liver tissue in the PV phase, gaining significant improvement by 30.6% (DPC). Even though the single ResSeg based on single-phase images performs not bad on the Cyst, HCC and METs, the segmentation performance can still be further improved by the exploitation of multi-phase information. Though our method can improve the segmentation of different types of tumors, a single segmentation network for five classes of tumors may cause the network balance among the classes and not fully explore its representation ability for every tumor class. In the future, we will take into account more class-specific information and may employ multi-task learning strategy to complete the segmentation and classification simultaneously. Furthermore, the segmentation of different types of tumors is regarded as a semantic segmentation task in the current work, namely the tumors are treated as a single class. A task of instance segmentation with specific label for each class is worthy exploring in the future work.

Table 11. Segmentation accuracy (DPC: %) of the proposed method regarding to different types of tumors. The data in parentheses shows the improvement of the PA-ResSeg over the single ResSeg.

| Network | Cyst | FNH | HCC | HEM | METs | Whole test set |
|---------|------|-----|-----|-----|------|----------------|
| Single ResSeg | 80.36 | 39.62 | 64.16 | 68.63 | 76.33 | 67.87 |
| PA-ResSeg | 85.70 | 73.13 | 71.89 | 74.63 | 79.64 | 77.26 |
| | (+5.34) | (+33.51) | (+7.73) | (+6.0) | (+3.31) | (+9.39) |

In order to better present the influence of BE-loss on marginal slices, Table 12 displays the comparison results of the models trained with the binary cross-entropy loss ($L_{ce}$) and the 3D BE-loss ($L_{be}$) regarding to marginal slices, middle slices and all slices.

Table 12. Comparison results of using different loss regarding to marginal slices, middle slices and all slices and the paired t-test results of the two models in terms of different metrics.

| | | | DPC | VOE | RVD | TPR | TNR | ACC |
|---|---|---|-----|-----|-----|-----|-----|-----|
| Marginal | MEAN | $L_{ce}$ | 0.6603 | 0.4640 | -0.1193 | 0.6395 | **0.9994** | 0.8194 |
| Slices | | $L_{be}$ | **0.6865** | **0.4391** | **0.0471** | **0.7500** | 0.9984 | **0.8742** |
| | $P$ | | 0.3664 | 0.4022 | 0.0311 | 0.0184 | 0.0025 | 0.019 |
| Middle Slices | MEAN | $L_{ce}$ | 0.7984 | 0.2069 | 0.5721 | 0.8656 | **0.9991** | 0.9323 |
| | | $L_{be}$ | **0.8104** | **0.1914** | **0.3831** | **0.9153** | 0.9990 | **0.9571** |
| | $P$ | | 0.4048 | 0.4607 | 0.0120 | 0.0029 | 0.0277 | 0.0029 |
| All Slices | MEAN | $L_{ce}$ | 0.7788 | 0.3315 | 0.2217 | 0.8073 | **0.9992** | 0.9032 |
| | | $L_{be}$ | **0.7828** | **0.3290** | **0.1173** | **0.8635** | 0.9989 | **0.9323** |
| | $P$ | | 0.7633 | 0.8763 | 0.0198 | 0.0023 | 0.0095 | 0.0023 |

A paired t-test was performed on the results of the two models to analyze the statistical difference (shown in Table 12). It is shown in the table that using BE-loss is able to further improve DPC and VOE of the results on both marginal slices and middle slices, whereas no significant difference was observed ($P > 0.05$) and can get better RVD, TPR and ACC with significant difference ($P < 0.05$). Results in Table 12 and Figure 8 show that the model trained using BE-loss is prone to over-segment tumors. Considering the actual situation that doctors tend to resect a larger area than the lesion to ensure the tumors removed completely as much as possible, an over-segmented tumor is better for subsequent application such as preoperative planning compared to an under-segmented result.

Although the proposed method gets a promising result, there are still some tumors not be properly segmented (shown in Figure 10). Figure 10(a) displays an example that have multiple lesions with different types and we can see from the figure that only one lesion of Cyst is segmented well. This may be caused by the lack of training data with multiple lesions and extreme data imbalance. In addition, the training of network is prone to be dominated by an easier task when there are several different lesions in a training sample. The class imbalance of data can also be observed from the detailed distribution of the MPCT-FLLs dataset in Table 1. To address this issue, on the one hand, we are going to extend the dataset in the future. On the other hand, we may further modify the loss function for handling the data imbalance to improve segmentation performance.

Image registration between phases is a prerequisite in multi-phase segmentation. We implement the registration by aligning two liver volumes using a simple yet effective B-spline based registration method. Hence, the segmentation performance of PA-ResSeg subjects to the accuracy of our registration step to some extent. Figure 11 presents an inaccurate registration example. It is observed that tumors between phases were not strictly aligned where the PV slice with a tumor was badly matched with an ART slice not containing a tumor. Accordingly, the tumor was not correctly segmented due to missing of complementary tumor information caused by inaccurate registration (see Figure 10(b)). Therefore, a much more effective registration algorithm is highly demanded. A model that does not require an aligned input pair of multi-phase images is also a possible direction to explore for segmentation of multi-phase CT images.

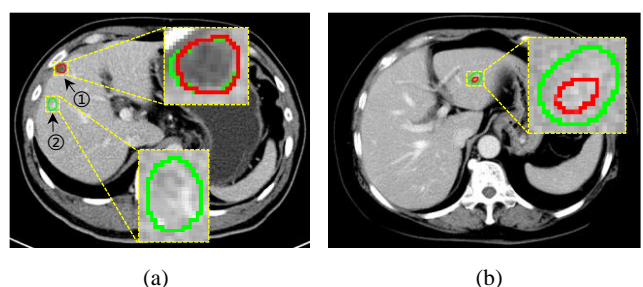

(a)                              (b)

Figure 10. Examples that are poorly segmented. (a) an example that have multiple tumors with different types. ① is a lesion of Cyst and ② is a lesion of HEM. (b) a poorly segmented example of FNH. Each yellow-square region is magnified. The green and red outlines represent ground truth and segmentation results, respectively.

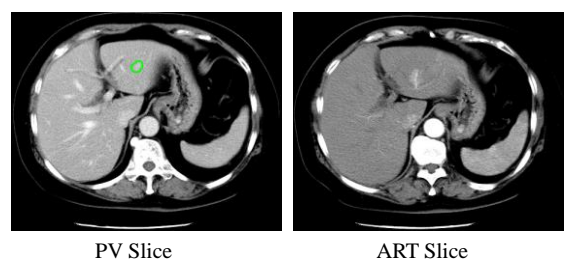

PV Slice                    ART Slice

Figure 11. An example of inaccurately registered image pair. The green outline denotes the tumor ROI.

In our experiments, we design the PA module to capture the inter-phase dependencies in the channel dimension. In the most attention-based approaches, usually both channel and spatial attention are employed, both of which are important. In our future work, we will extend the PA module to explore the spatial inter-phase correlations, not only the channel-wise dependencies, expecting to generate more representative multi-phase information for liver tumor segmentation.

## 5. CONCLUSIONS

In this work, we present a phase attention residual network (PA-ResSeg) for accurate liver tumor segmentation based on multi-phase CT images. A PA-based MSF strategy is exploited to effectively probe the complementary multi-phase information, where the PA blocks are designed to learn more representative multi-phase features by capturing the intra-phase dependency and the inter-phase dependency and the MSF structure can incorporate multi-scale features from multiple phases. Furthermore, we propose a 3D boundary-enhanced loss to handle the situation of poor segmentation of boundaries and the marginal slices. The effectiveness of each component is validated through an ablation study and the experiments on the MPCT-FLLs dataset and the MPCT-HCC dataset demonstrate the superiority and robustness of our proposed method.

## ACKNOWLEDGEMENTS

This work was supported in part by China Postdoctoral Science Foundation (No. 2020TQ0293), in part by Major Scientific Project of Zhejiang Lab (No. 2020ND8AD01), in part by the Grant-in Aid for Scientific Research from the Japanese Ministry for Education, Science, Culture and Sports (MEXT) under the Grant No. 20KK0234, No. 18H03267 and No. 20K21821, and in part by Grant No. HTKJ2019KL502005.

## CONFLICT OF INTEREST STATEMENT

The authors have no relevant conflicts of interest to disclose.